\documentclass[aps,eqsecnum,prd,twocolumn]{revtex4}
\usepackage{graphics,graphicx}
\usepackage{amsmath}
\usepackage{amssymb,latexsym,mathrsfs}
\usepackage{hyperref}

\def\bea{\begin{eqnarray}}
\def\eea{\end{eqnarray}}
\def\ba{\begin{array}}
\def\ea{\end{array}}

\def\beq{\begin{equation}}
\def\eeq{\end{equation}}

\begin{document}

\title{Temperature dependent Zeno time for a two level atom tunneling through a thermal magnetic barrier in the framework of weak measurement}

\author{Samyadeb Bhattacharya$^{1}$ \footnote{sbh.phys@gmail.com}, Sisir Roy$^{2} $ \footnote{sisir@isical.ac.in}}
\affiliation{$^{1,2}$Physics and Applied Mathematics Unit, Indian Statistical Institute, 203 B.T. Road, Kolkata 700 108, India \\}

\vspace{2cm}
\begin{abstract}

\vspace{1cm}

\noindent The Zeno time has been calculated for a metastable two level atom tunneling through a interacting thermal magnetic field. The process of weak measurement has been utilized for the the estimation of the timescale. The temperature dependence of the Zeno time has been shown. From the calculation it is evident that the Zeno time decreases with the increase of temperature. Moreover, the result restricts the Zeno time to a maximum limiting value, irrespective of how frequent is the measurement process.

\vspace{2cm}


\vspace{1cm}

\end{abstract}

\vspace{1cm}

\maketitle

\section{Introduction}

Quantum Zeno effect is the phenomena of inhibition of transition between quantum states by the process of frequent non-selective measurement \cite{1,2,3,4,5,6,7}. The short-time behavior of the non-decay probability of any unstable quantum particle is shown to be quadratic, not exponential \cite{8}. Mishra and Sudarshan \cite{1} showed that this curious behavior when combined with the theory of quantum measurement can lead to a very surprising conclusion that frequent non selective observation slows down or even freezes the decay process. Under frequent observations (measurements), unstable quantum particle stops to decay. This effect has been successfully demonstrated previously by many experimental observations \cite{9,10,11,12,13,14,15,16}. Zeno dynamics has got remarkable applications in the fields of decoherence control \cite{17}, state purification \cite{18}, implementation of quantum gates \cite{19}, entanglement protection \cite{20} etc. These processes are instrumental for building quantum memory devices, which is the main component in construction of quantum computers. Ion trapping mechanism seems to be a very good candidate for making such memory devices. There also, preserving quantum coherence is a very challenging task, because the quantum entity constantly undergoes environmental interactions. Ion trapping systems are excellent candidate for the measurement of long coherent processes in atomic systems. It is very much possible to use long coherent pulses for the coupling of two atomic levels and then study their dynamics. As we have mentioned earlier, many theoretical and experimental studies show that quantum Zeno effect is a nice way to control decoherence. In this work, our aim is to find an expression for the time scale for Zeno effect, for a two level atom tunneling through a barrier of thermal magnetic field. Whenever frequent non-selective measurement dominates the time evolution of the state, the system is forced to evolve in a subspace of the total Hilbert space, which is called "Zeno subspace". Probability leakage is not possible between these invariant Zeno subspaces. So each of these subspaces can be considered as some reduced isolated system. Due to their isolated nature, the process of decoherence can be halted within these Zeno subspaces. Now, if the environmental interaction is very strong, extreme decoherence may not allow the Zeno subspaces to sustain. So Zeno effect, characterized by it's corresponding time scale (Zeno time), gives a certain lower limit to decoherence, below which it is uncontrollable. Our main investigation is to find how this time scale is dependent on the temperature of the surroundings. In order to do that, we will use the framework of weak measurement \cite{21,22,23,24}. The weak value of a quantum mechanical observable is the statistical result of a standard measurement procedure performed upon a pre selected and post selected (PPS) ensemble of quantum systems when the interaction between the measurement apparatus and each system is sufficiently weak. Unlike the standard strong measurement of a quantum observable, weak measurement does not appreciably disturb the quantum system and yields the weak value as the measured value of the observable. In this type of measurement scheme, the interaction between the system and the measuring device (in our case the thermal magnetic field) is made very small. This is a good way to minimize the environmental interaction, which is in turn helpful for reducing the effect of decoherence. Another important feature of weak measurement process is that, here we take an ensemble average of numerous observations over the pre-selected and post-selected states, because one single measurement interaction cannot bring out enough information about the system. Since the Zeno dynamics is initiated by frequent observations, an ensemble average over many such observations is necessary to observe the dynamics over a finite period of time. So weak measurement scheme is also very much compatible with the Zeno type measurement procedure. In the next section, we will discuss the formulation of Zeno time in the usual algebraic procedure and then find it's value in the scheme of weak measurement. In section III, we will then concentrate on the master equation for a two level atom tunneling through a thermal magnetic barrier and find the decay parameter, which will be essential for the final expression of the weak Zeno time. After that we will conclude with possible implications.

\section{Weak value of Zeno time}

Quantum Zeno Effect (QZE) can be theoretically presented in a very simple way with the consideration of the short time behavior of the decaying quantum state \cite{25}. If $|\psi\rangle$ is the decaying quantum state at initial time $t=0$ and $H_s$ is the system Hamiltonian, then the state vector  of the system at final time $t$ can be represented as $e^{-\frac{iH_st}{\hbar}}|\psi\rangle$. Then the survival probability is given by
\beq\label{z1}
P(t)=|\langle \psi|e^{-\frac{iH_s t}{\hbar}}|\psi\rangle|^2
\eeq
For small time interval $t$, a power series expansion of the time evolution upto $2^{nd}$ order
\beq\label{z2}
\exp\left(-\frac{iH_s t}{\hbar}\right)=1-\frac{iH_s t}{\hbar}-\frac{1}{2}\frac{H_s^2}{\hbar^2} t^2
\eeq
So then the modified survival probability is given by
\beq\label{z3}
P(t)=|\langle \psi|e^{-\frac{iH_s t}{\hbar}}|\psi\rangle|^2\approx \left[1-\frac{(\Delta H_s)^2}{\hbar^2}t^2\right]
\eeq
where
\beq\label{z4}
(\Delta H_s)^2=\langle \psi|H_s^2|\psi\rangle-\langle \psi|H_s|\psi\rangle^2
\eeq
is the uncertainty in system energy measurement. There are many quantum mechanical states having survival probability which appears to be decreasing exponentially on ordinary time scales. The quadratic time dependence of \ref{z1} is naturally inconsistent with those decaying states and implies that in such cases \ref{z1} holds only for very short time intervals. Consider the survival probability $P(t)$, where the interval $[0,\tau]$ is interrupted by $n$ frequent non-selective measurements done on equal interval at times $\tau/n,2\tau/n,....\tau$ . In an ideal scenario, these measurements are nothing but instantaneous projections. The initial state $|\psi\rangle$ of the concerning system is of course an eigenstate of the measurement operator. Then the survival probability can be given by
\beq\label{z5}
P(t)\approx \left[1-\frac{(\Delta H_s)^2}{\hbar^2} (\tau/n)^2\right]^n
\eeq
which approaches 1 as $n\rightarrow \infty$. The Quantum Zeno Effect (QZE) can be observed as long as the quantum system displays the behavior shown in \ref{z5}. From \ref{z3} we can construct a time scale as
\beq\label{z6}
\tau^Z=\frac{\hbar}{\Delta H_s}
\eeq
which is called the "Zeno time" or the timescale within which Zeno effect can be observed. From \ref{z3} we can easily infer that as long as the time interval is shorter than the Zeno time, the system is freezed in the initial state. If the interval between consecutive measurements on the system is smaller than $\tau^Z$, the dynamics of decay significantly slows down or even can be asymptotically halted. Now if $\tau^M$ is the measurement time, then \ref{z5} can be written as
\beq\label{z7}
P(t)\approx \left[1-\left(\frac{\tau^M}{\tau^Z}\right)^2\right]^{\tau/\tau^M}
\eeq
It follows that
\beq\label{z8}
P(\tau)\approx \exp\left[-\tau\tau^M/(\tau^Z)^2\right]
\eeq
The only condition to go from \ref{z7} to \ref{z8} is $\tau^M \ll \tau^Z$ ( there is no restriction on $\tau$ ). So if the lifetime of the decaying state is $\tau^L$, then $P(\tau)=e^{-\tau/\tau^L}$, we can define
\beq\label{z9}
\tau^Z\approx \sqrt{\tau^L \tau^M}
\eeq
But in practical situation, Nakazato et.al \cite{25a} showed that the $N\rightarrow\infty$ limit is only a mathematical limit not realizable in experimental situation. They have experimentally showed that in real situation $N$ is actually a finite number which is not very large. On a microscopic scale, a measurement process, as a physical process takes place in a considerable duration of time, although in a macroscopic scale it can be considered instantaneous.  \\
Based on the theoretical background we just discussed, we are now going to introduce the framework of weak measurement, which we are going to use to calculate the Zeno time. The reason behind using this particular measurement framework is that in our consideration the interaction between the thermal electromagnetic field and the system is sufficiently weak. So one single measurement or interaction on the system does not give any significant result; or in other words does not disturb the system in a considerable way. So numerous such interactions are taken into account to get an ensemble average for some quantum observable. In this work we will use the method originally developed by Davies \cite{26}. In that work, Devies used the time dependent weak value of projection operator to determine a generalized survival probability of a decaying two state system. \\
The time evolution of the state is considered to be
\beq\label{z10}
|\psi(t)\rangle= U(t-t_0)|\psi(t_0)\rangle
\eeq
where the time evolution operator
\beq\label{z11}
U(t-t_0)=e^{-iH_s(t-t_0)/\hbar}
\eeq
The expression of the time dependent weak value of a certain operator $A$ pre -selected at time $t_i$ and post selected at $t_f$
\beq\label{z12}
A_w= \frac{\langle\psi_f|U^{\dagger}(t-t_f)AU(t-t_i)|\psi_i\rangle}{\langle\psi_f|U^{\dagger}(t-t_f)U(t-t_i)|\psi_i\rangle}
\eeq
If we now consider that the two level atom is in a barrier region of external magnetic field in the z direction. Then the system Hamiltonian can be represented as
\beq\label{z13}
H_s=\frac{1}{2}\hbar \Omega \sigma_z
\eeq
where $\sigma_z$ is the Pauli spin matrix in the z direction and $\Omega$ is the Rabi oscillation frequency. So the time evolution operator looks like
\beq\label{z14}
U(t)= \left (  \begin{array}{ll}
                 e^{i\Omega t/2} & 0\\
                 0 & e^{-i\Omega t/2}
                 \end{array} \right)
\eeq
If the initial pre-selected state at $t_i$ is x polarized then
\beq\label{z15}
|\psi_i\rangle= \frac{1}{\sqrt{2}}\left (\begin{array}{ll}
                         1\\
                         1
                 \end{array} \right)
\eeq
and the associated projection operator
\beq\label{z16}
P_{+}= \frac{1}{\sqrt{2}}\left (  \begin{array}{ll}
                 1 & 1\\
                 1 & 1
                 \end{array} \right)
\eeq
In case of the decay of any metastable state, the system of two level atom is considered to be coupled to $2N$ number of environmental bath modes, which are initially in their ground states. Because of the presence of the interaction with these bath modes, the system loses energy to them. For simplicity we consider that any arbitrary state $E_n$ satisfies the relation
\beq\label{z17}
E_n-E_0=n\Delta E,~~~~~-N\leq n\leq N
\eeq
with the assumption that the reference atom is equally coupled to all the bath modes. Following Davies \cite{26} we find that
\beq\label{z18}
a_0(t)= e^{-\Gamma (t-t_i)}
\eeq
where $a_0$ is the amplitude of the pre-selected initial state and $\Gamma$ is the population inversion decay parameter for the two level metastable state. The time evolution operator for the decaying state $U(t)$ is a $(2N+1)\times(2N+1)$ dimensional matrix with the components $U_{ij}$, where
\beq\label{z19}
U_{00}=e^{-\Gamma t}
\eeq
with the limit $\Delta E\rightarrow 0$. Under the relation $U^{\dagger}(t)=U(-t)$, using \ref{z12} for the projection operator $P_{+}$, we get the weak value for the same pre and post selected state \cite{26,27}
\beq\label{z20}
P_w=e^{-\Gamma(t-t_i)} \left[\frac{1-e^{-\Gamma(t_f-t)}}{1-e^{-\Gamma(t_f-t_i)}}\right]
\eeq
This is the time interval between two successive interactions between the system and the field. From \ref{z20} we can clearly see that
\beq\label{z21}
\begin{array}{ll}
P_w=1~~~~~ for~~t=t_i\\
~~~~=0~~~~for~~t=t_f
\end{array}
\eeq
So \ref{z20} gives the generalized weak decay law. At initial time $t_i$ the particle is pre-selected in the higher state and it is post-selected at $t_f$ in the same state, when that state is decayed due to the interaction with the environment. Now let us compare this weak survival probability with the expression of probability given by \ref{z7}. There we find that if $\tau_M=0$, the probability $P$ is 1 and for $\tau_M=\tau_z$, the probability $P$ is 0. The time integral in the span of the measurement time $\tau_M$ over the generalized survival probability gives us the decay time ($\tau_L$), within which time scale the system can freezed in it's initial state by repetitive measurements . Now time integrating \ref{z20} over the interval $t_i$ to $t_f$ with the approximation of small dissipation: $(t_f-t_i)\ll1/\Gamma$, we get
\beq\label{z22}
\tau_{L}=\frac{1}{\Gamma+2/(t_f-t_i)}
\eeq
The total time interval $t_f-t_i$ can be taken as equal to $N\tau_M$, where $N$ is a large but finite number. Now using the approximate expression of Zeno time from \ref{z9}, we get

\beq\label{z23}
\tau_z=\sqrt{\frac{\tau_M}{\Gamma+2/N\tau_M}}
\eeq

The  population inversion decay parameter $\Gamma$ and the measurement or interaction time will depend on the nature of interaction with the electromagnetic field. Now in the next section it is our aim to determine the exact form of those mentioned parameters from the master equation of the two atom in a thermal magnetic field.

\section{Temperature dependence of Zeno time}

In this section we will concentrate on the time evolution of the two level atom coupled to a thermal magnetic field to determine the decay parameter, using their corresponding master equation \cite{28}. The total Hamiltonian of system plus reservoir can be described by
\beq\label{t1}
H_T=H_s+H_f+H_i
\eeq
where $H_s$ is the system Hamiltonian described by \ref{z13}. $H_f$ and $H_i$ are respectively the field and interaction Hamiltonian given by
\beq\label{t2}
H_f=\sum_n \hbar\omega_n \hat{a}^{\dagger}\hat{a}
\eeq

\beq\label{t3}
H_i(t)=g\left(\sigma_{+}e^{i\Omega t}+\sigma_{-}e^{-i\Omega t}\right)B(t)
\eeq
where

\beq\label{t4}
B(t)=\exp\left(-\frac{H_f}{i\hbar}t\right)B\exp\left(\frac{H_f}{i\hbar}t\right)
\eeq
This is the magnetic field operator in the interaction picture containing a wide range of frequencies. Among those frequencies, we are only concerned with the ones almost in resonance with $\pm \Omega$. Under this approximation, the correlation function is described as \cite{28}

\beq\label{t5}
e^{i\Omega(t-t')}\langle B(t)B(t') \rangle\sim 4\hbar \Omega (N(\Omega)+1)\delta(t-t')
\eeq

Similarly

\beq\label{t6}
e^{-i\Omega(t-t')}\langle B(t)B(t') \rangle\sim 4\hbar \Omega N(\Omega)\delta(t-t')
\eeq

$N(\Omega)$ is the Planck function given by

\beq\label{t7}
N(\Omega)=\frac{1}{\exp\left(\frac{\hbar \Omega}{KT}\right)-1}
\eeq

Now the master equation for the the reduced density operator $\widetilde{\rho_i}$ is given by \cite{28}

\beq\label{t8}
\begin{array}{ll}
\frac{d\widetilde{\rho_i}}{dt}=\frac{2g^2\Omega}{\hbar} (N(\Omega)+1)\left[\sigma_{-}\widetilde{\rho_i}\sigma_{+}-\sigma_{+}\sigma_{-}\widetilde{\rho_i}-\widetilde{\rho_i}\sigma_{+}\sigma_{-}\right]\\
~~~~~~ +\frac{2g^2\Omega}{\hbar}N(\Omega)\left[\sigma_{+}\widetilde{\rho_i}\sigma_{-}-\sigma_{-}\sigma_{+}\widetilde{\rho_i}-\widetilde{\rho_i}\sigma_{-}\sigma_{+}\right]

\end{array}
\eeq

Considering a slightly more intricate situation, where we take a complex electromagnetic field

\beq\label{t9}
B(t)\rightarrow B(t)+\Lambda e^{i\Omega t}+\Lambda^{*} e^{-i\Omega t}
\eeq

Now the modified master equation is given as

\beq\label{t10}
\begin{array}{ll}
\frac{d\widetilde{\rho_i}}{dt}=\frac{2g^2\Omega}{\hbar} (N(\Omega)+1)\left[\sigma_{-}\widetilde{\rho_i}\sigma_{+}-\sigma_{+}\sigma_{-}\widetilde{\rho_i}-\widetilde{\rho_i}\sigma_{+}\sigma_{-}\right]\\
~~~~~~ +\frac{2g^2\Omega}{\hbar}N(\Omega)\left[\sigma_{+}\widetilde{\rho_i}\sigma_{-}-\sigma_{-}\sigma_{+}\widetilde{\rho_i}-\widetilde{\rho_i}\sigma_{-}\sigma_{+}\right]\\
~~~~~~-\frac{ig}{\hbar}\left[\left(\Lambda \sigma_{+}+\Lambda^{*}\sigma_{-}\right),\widetilde{\rho_i}\right]

\end{array}
\eeq

From this master equation the time evolution of the expectation value of the Pauli spin operators can be given as

\beq\label{t11}
\begin{array}{ll}
\frac{d\langle\sigma_{+}\rangle}{dt}=-\frac{2g^2 \Omega}{\hbar}(2N(\Omega)+1)\langle\sigma_{+}\rangle-\frac{ig}{\hbar}\Lambda^{*}\sigma_z \\
\frac{d\langle\sigma_{-}\rangle}{dt}=-\frac{2g^2 \Omega}{\hbar}(2N(\Omega)+1)\langle\sigma_{-}\rangle+\frac{ig}{\hbar}\Lambda\sigma_z \\
\frac{d\langle\sigma_z\rangle}{dt}=-\frac{4g^2 \Omega}{\hbar}(2N(\Omega)+1)\langle\sigma_z\rangle-\frac{4g^2 \Omega}{\hbar}\\
~~~~~~~~~~-\frac{i}{2}\frac{g}{\hbar}(\Lambda\langle\sigma_{+}\rangle-\Lambda^{*}\langle\sigma_{-}\rangle)
\end{array}
\eeq

From \ref{t11} we can see that the system state denoted by $\langle\sigma_z\rangle$ decays at the rate

\beq\label{t12}
\Gamma=\frac{4g^2 \Omega}{\hbar}(2N(\Omega)+1)=\frac{4g^2 \Omega}{\hbar}\coth^2\left(\frac{\hbar\Omega}{2KT}\right)
\eeq

\ref{t12} represents the rate of population inversion, which we take as the decay parameter. Again $\tau_M$ is the time interval between two successive measurement interactions. Now the amplitude of the magnetic field varies with it's characteristic frequencies, of which only the resonant frequency $\Omega$ is important here. After the time period, which equals to the inverse of this characteristic frequency, the magnetic field becomes maximum. So this time scale can be argued to be the time interval between two successive maximum measurement interactions. So we take $\tau_M$ as the inverse of this characteristic frequency $\Omega$. Now we take total time interval $t_f-t_i=N\tau_M$, where $N$ is a large but finite number. So using these values of $\Gamma$ and $\tau_M$ in \ref{z23}, we get the expression of Zeno time as

\beq\label{t13}
\tau_z= \sqrt{\frac{N}{2}}\frac{1}{\Omega}\left(1+\frac{2g^2N}{\hbar}\coth^2\left(\frac{\hbar\Omega}{2KT}\right)\right)^{-1/2}
\eeq

The variation of Zeno time with temperature is shown in the FIG 1.

\begin{figure}[htb]
{\centerline{\includegraphics[width=7cm, height=5cm] {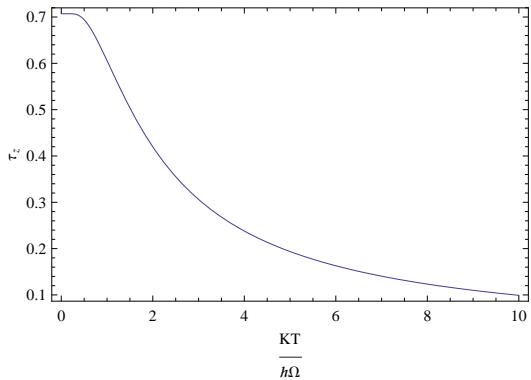}}}
\caption{$\tau_z$ vs. $\frac{KT}{\hbar\Omega}$ }
\label{figVr}
\end{figure}

From the figure it can be clearly seen that the Zeno time decreases with increasing temperature. This is very much compatible with experimental situation. Because with increasing temperature the environmental interaction increases, making the decay process stronger. For that reason, it becomes harder to make Zeno effect realizable with increasing temperature. The maximum Zeno time is at zero temperature given by the expression

\beq\label{t14}
\tau_z^0=\sqrt{\frac{N}{2}}\frac{1}{\Omega}\left(1+\frac{2g^2N}{\hbar}\right)^{-1/2}
\eeq

Now if $N$ can be increased to a very large number, making the measurement procedure quasi-continuous, then the Zeno time at zero temperature becomes

\beq\label{t15}
\tau_z^0|_{N\rightarrow \infty}= \frac{1}{2\Omega}\frac{\sqrt{\hbar}}{g}
\eeq
which shows that even if we make the measurement quasi-continuous, Zeno time will only increase to a certain limiting value.
\section{Conclusion}

In this paper, we have formulated the expression of Zeno time as a function of temperature and other parameters of system-bath interaction, for a two level atom tunneling through a thermal magnetic field. We have used the procedure of weak measurement for our calculation. The reason behind using this particular measurement framework is that in this type of measurement scheme, the interaction between the system and the measuring device (in this case thermal magnetic field) is made very small. This can be an useful way to restrain the environmental interaction, which is in turn helpful to control the decohering process. Other important feature of weak measurement process which should be mentioned here is that, in this measurement scheme an ensemble average of numerous observations is taken over the pre-selected and post-selected states. Here one single measurement interaction is not sufficient to bring out necessary information about the system. Since the Zeno process is initiated by frequent non-selective measurements, an ensemble average over many such measurement interactions is necessary to observe the dynamics over a finite period of time. So scheme of weak measurement is also compatible with the Zeno type measurement process. In this work, our calculation shows that the Zeno time decreases with the increase of temperature, which is compatible with practical situations. Because with the increase of temperature the system-bath interaction increases. So the process of decoherence due to environmental dissipative interaction gets stronger. The essence of Zeno dynamics is that due to non-selective frequent measurements the total Hilbert space reduces to a quasi-isolated reduced subspace, within which the decay dynamics can be stopped or at least considerably slowed down. Now if the process of decoherence gets stronger, then isolating the system in the reduced Zeno subspace gets much more harder. So with the increase of temperature as the decoherence process gets stronger compared to the Zeno process, the Zeno time decreases. Again from \ref{t15} we can see that even quasi-continuous measurement cannot make the the Zeno time infinitely large. The Zeno time can only be increased to a limiting value. So our result also imposes a restriction over the method of non-selective frequent measurement procedure, giving a limiting maximum value of Zeno timescale irrespective of how frequent is  the measurement process. The reason behind this limitation is that even at zero temperature, the coupling between the system and environment exists. So even at zero temperature, the reduced Zeno subspaces are vulnarable to the environment induced decoherence process, which gives a restriction over the Zeno dynamics.

\end{document}